\documentclass[12pt]{iopart}

\usepackage{graphicx}
\usepackage{amssymb}
\usepackage{wasysym}
\usepackage{subfigure} 
\usepackage{color}

\begin{document}

\title{Finite-size scaling method for the Berezinskii-Kosterlitz-Thouless transition}

\author{Yun-Da Hsieh$^1$, Ying-Jer Kao$^1$ and Anders W Sandvik$^2$}
\vskip2mm

\address{$^1$Center of Theoretical Sciences and Department of Physics,\hfill\break 
\null~~National Taiwan University, No. 1, Sec. 4, Roosevelt Road, Taipei 10607, Taiwan \ead{yjkao@phys.ntu.edu.tw}}
\vskip2mm

\address{$^2$ Department of Physics, Boston University, 590 Commonwealth Avenue,\hfill\break  
\null~~~Boston, Massachusetts 02215, USA
\ead{sandvik@buphy.bu.edu}}

\begin{abstract}
We test an improved finite-size scaling method for reliably extracting the critical temperature $T_{\rm BKT}$ of a Berezinskii-Kosterlitz-Thouless 
(BKT) transition. Using known single-parameter logarithmic corrections to the spin stiffness $\rho_s$ at $T_{\rm BKT}$ in combination with the
Kosterlitz-Nelson relation between the transition temperature and the stiffness, $\rho_s(T_{\rm BKT})=2T_{\rm BKT}/\pi$, we define a size dependent 
transition temperature $T_{\rm BKT}(L_1,L_2)$ based on a pair of system sizes $L_1,L_2$, e.g., $L_2=2L_1$. We use Monte Carlo data for the standard 
two-dimensional classical XY model to demonstrate that this quantity is well behaved and can be reliably extrapolated to the thermodynamic limit 
using the next expected logarithmic correction beyond the ones included in defining $T_{\rm BKT}(L_1,L_2)$. For the Monte Carlo calculations we 
use GPU (graphical processing unit) computing to obtain high-precision data for $L$ up to $512$. We find that the sub-leading logarithmic corrections 
have significant effects on the extrapolation. Our result $T_{\rm BKT}=0.8935(1)$ is several error bars above the previously best estimates of the 
transition temperature; $T_{\rm BKT} \approx 0.8929$. If only the leading log-correction is used, the result is, however, consistent with the
lower value, suggesting that previous works have underestimated $T_{\rm BKT}$ because of neglect of sub-leading logarithms. Our method is easy to 
implement in practice and should be applicable to generic BKT transitions.
\end{abstract}


\maketitle

\section{Introduction}
\label{sec:intro}

The Berezinskii-Kosterlitz-Thouless (BKT) transition \cite{berezinskii72,kosterlitz73,kosterlitz74} is very well understood in terms of its physical 
mechanism of vortex-antivortex unbinding. The field-theoretical formulation of this two-dimensional (2D) problem of an U($1$) symmetric order parameter
gives a rigorous quantitative characterization of the transition into the critical (``quasi-ordered'') state obtaining below $T_{\rm BKT}$. There are
also exactly solvable models with BKT transitions \cite{beijeren77,nienhuis82}. Despite the detailed theoretical understanding of the BKT transition, 
analyzing numerical data from Monte Carlo (MC) simulations (or other numerical techniques) of the transition on finite lattices is still challenging 
\cite{schultka94,schultka95,harada97,tomita02,melko04,hasenbusch05a,hasenbusch05b,hasenbusch08,carrasquilla12,komura12,iaconis12,pelissetto13}, 
because of the presence of logarithmic finite-size corrections \cite{weber87,kenna97}. It has been an ongoing quest to find detailed forms of these 
logarithmic corrections to high order \cite{hasenbusch05a,hasenbusch05b,hasenbusch08,iaconis12,pelissetto13} and to device fitting procedures to take 
them properly into account when analyzing finite-size data.

We here formulate an improved procedure for extracting the BKT transition temperature $T_{\rm BKT}$ in the thermodynamic limit using a finite-size definition 
$T^*(L_1,L_2)$ of $T_{\rm BKT}$ based on the spin stiffness (helicity modulus) $\rho_s$ for a pair of system sizes $L_1,L_2$. We wish to incorporate 
from the outset the Nelson-Kosterlitz (NK) criterion \cite{nelson77} for the discontinuity of the stiffness in the thermodynamic limit,
\begin{equation}
\rho_s(T_{\rm BKT}) = \frac{2T_{\rm BKT}}{\pi}.
\label{rhojump}
\end{equation}
In order to use this condition also for finite size, we write the stiffness as a function of the system size $L$ on an $L\times L$ (or some
non-square shape) lattice as
\begin{equation}
\rho_s(T_{\rm BKT},L) = \rho_s(T_{\rm BKT},\infty)F(L),
\label{rhotstar}
\end{equation}
where $F(L)$ represents the finite-size correction, $F(L) \to 1$ when $L \to \infty$. We next define a temperature $T^*(L_1,L_2)$ 
for a pair of system sizes $L_1,L_2$ such that
\begin{equation}
\frac{\rho_s(T^*,L_1)}{F(L_1)} = \frac{\rho_s(T^*,L_2)}{F(L_2)} = \frac{2T^*}{\pi}.
\label{rhol1l2}
\end{equation}
The reason why the two equalities can hold simultaneously is that the correction $F(L)$ contains a single unknown constant, which can be
regarded as a fitting parameter, chosen such that both equalities are satisfied at a unique value of the temperature $T=T^*$. Since $F(L) \to 1$ 
when $L\to \infty$ the NK relationship (\ref{rhojump}) holds in this limit and $T^* \to T_{\rm BKT}$. This procedure of taking advantage of the
NK relationship is more elaborate than the “curve crossing” method often used when analyzing dimensionless quantities at conventional phase 
transitions \cite{sandvik10}, but it is still rather easy to apply. More standard curve crossing methods have also been used when analyzing 
the BKT transition \cite{carrasquilla12} and some attempts to incorporate the NK criterion along the lines above have also been made 
\cite{hasenbusch05b}. We here go to higher order than previously and also include further logarithmic corrections when extrapolating
$T^*$ to the thermodynamic limit.

Using the standard 2D classical XY model, we systematically investigate the finite-size dependence of $T^*$ when increasingly sophisticated 
forms of the correction $F(L)$ are used. We find that it is crucial to use the most complete available form of the logarithmic corrections. 
We find results for $T_{\rm BKT}$ comparable to those in several recent works if only the leading logarithmic corrections are taken 
into account in $F(L)$ and a naive power-law finite-size extrapolation of $T^*$ is used. However, when all known logarithmic corrections are
taken into account properly we obtain a significantly higher $T_{\rm BKT}$. Our final estimate is $T_{\rm BKT}=0.8935(1)$, while the previously 
best estimates are clustered around $0.8929$ \cite{hasenbusch05a,hasenbusch08,komura12}.

The outline of the rest of the paper is as follows.
In Sec.~\ref{sec:corr} we discuss the details of the known corrections to the spin stiffness and how we take these into account
in our fitting procedures. We discuss the MC techniques in Sec.~\ref{sec:mc} and the results in Sec.~\ref{sec:res}. We conclude
with a brief summary and discussion in Sec.~\ref{sec:disc}.

\section{Logarithmic corrections and stiffness renormalization factors}
\label{sec:corr}

We begin here by discussing two different forms of the multiplicative correction $F(L)$ in Eq.~(\ref{rhol1l2}), based on leading and 
higher-order logarithmic forms. We also discuss the form of the leading remaining corrections not included in $F$, and renormalization
factors entering for stiffness estimators used in MC simulations. With all these results from previous works collected, we discuss
our method to use them in practice together with the NK relationship (\ref{rhojump}).

\subsection{Logarithmic corrections}

\begin{figure}
\begin{center}
\includegraphics[width=9cm, clip]{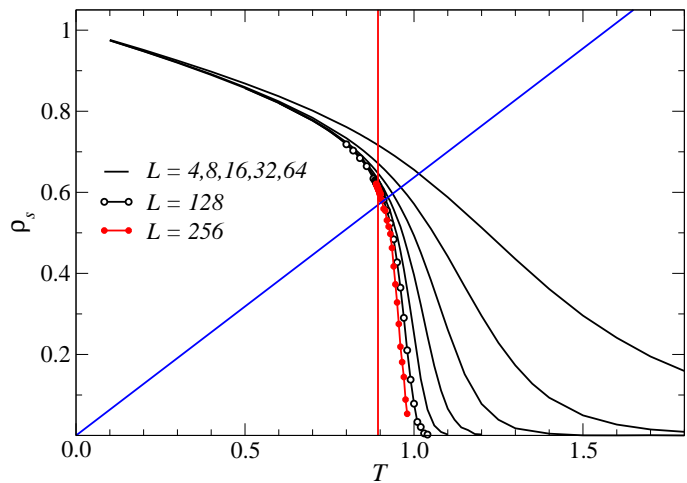}
\end{center}
\vskip-4mm
\caption{MC results for the spin stiffness of the 2D classical XY model for several lattice sizes of the form $L=2^n$. A discontinuity develops 
at $T_{\rm BKT}$ when $L \to \infty$, at a point satisfying the NK relation, Eq.~(\ref{rhojump}), indicated here by the line $\rho_s=2T/\pi$. The 
vertical line is the actual transition temperature $T_{\rm BKT}\approx 0.8935$ (as determined in this paper) of the model. Thus, the intersection 
of the two lines is at $\rho_s(T_{\rm BKT})$.} 
\label{rhos}
\end{figure}

Weber and Minnhagen (WM) derived the following logarithmic finite-size correction to the spin stiffness exactly at the
transition temperature \cite{weber87};
\begin{equation}
\rho_s(T_{\rm BKT},L) = \rho_s(T_{\rm BKT},\infty)\left ( 1 + \frac{1}{2\ln(L) + C} \right ),
\label{rhoslogcorr}
\end{equation}
where $C$ is an unknown constant (which turns out to not be a constant but is size-dependent, as discussed below) which depends on the microscopic 
details of the system under study. We illustrate the slow convergence in Fig.~\ref{rhos} by plotting raw MC results for $\rho_s$ for the classical 
2D XY model (we will describe the calculations below in Sec.~\ref{sec:mc}) for different system sizes versus the temperature.

Higher-order corrections are now known from more detailed studies of the renormalization-group flows around the BKT 
transition \cite{hasenbusch05a,hasenbusch05b,hasenbusch08,pelissetto13}. The finite-size to infinite-size stiffness ratio can 
be written in the form
\begin{equation}
\frac{\rho_s(T_{\rm BKT},L)}{\rho_s(T_{\rm BKT},\infty)}=1 + \frac{1}{2\ln(L) + C+\ln[C/2 + \ln(L)]} + \frac{a}{\ln^2(L)}+\ldots,
\label{rhoslogcorr1}
\end{equation}
where $a$ is another unknown constant. In principle, the additional term $\ln[C/2 + \ln(L)]$ in the denominator beyond the WM form
can also be taken into account by expanding to leading order for large enough $L$. This can be combined with the $a/\ln^2(L)$ term to give a 
correction of the form $\propto \ln[\ln(L)]/\ln^2(L)$ to the WM form. We will test both these approaches when fitting data.

\subsection{Stiffness renormalization}
\label{renormalization}

An interesting complication for finite-lattice calculations of $\rho_s$ was noted some time ago by Prokof'ev and Svistunov \cite{prokofev00}: For a 
system on a torus (i.e., with periodic boundary conditions in both directions of the 2D square lattice), the stiffness measured in the standard
way in simulations [in the case of the classical XY model using Eq.~(\ref{rhosestimator2cl}) in Sec.~\ref{sec:mc}] does not give $\rho_s$ exactly.
It is affected by  a normalization factor depending on the aspect ratio $R=L_x/L_y$ of an $L_x\times L_y$ lattice. This is because the derivation of 
(\ref{rhosestimator2cl}) based on imposing a twist (see, e.g., Ref.~\cite{sandvik10}) assumes that there is no net flux field threading the torus
apart from the externally twist-imposed one, while in fact such ``field quanta'' are thermally excited in the the torus at any finite temperature, and 
they renormalize the stiffness in two dimensions (but there is no such effect in three dimensions). In the limit $L_x\to \infty, L_y\to \infty$, 
the stiffness measured in MC simulations according to (\ref{rhosestimator2cl}) in the $x$ and $y$ direction is related to the stiffness $\rho_s$ 
appearing in the BKT action and in Eqs.~(\ref{rhojump}) and (\ref{rhoslogcorr}) according to;
\begin{equation}
\rho^{\rm MC}_x = f_x(R)\rho_s,~~~~~\rho^{\rm MC}_y = f_y(R)\rho_s,
\label{rhonorm}
\end{equation}
where $f_x \not= f_y$ unless $R=1$ and $f_x \to 1$, $f_y\to 0$ when $R\to \infty$.

Fortunately, the renormalization factors $f_x,f_y$ due to the thermally excited flux quanta can be easily computed numerically (and in a special case 
analytically in terms of Ramanujan's $\Theta$-function \cite{melko04}); a list for selected aspect ratios is given in Ref.~\cite{melko04}. Here we will use 
$R=1$, for which $f_x=f_y=f=0.99982471$ \cite{prokofev00}. As previously noted in Ref.~\cite{melko04}, Monte Carlo calculations of $T_{\rm BKT}$ have in
the past typically not reached the level of precision where this factor would play any role (for $R=1$, which is normally used), but in high-precision 
calculations the renormalization should be included in order to avoid a systematical error. Our calculations here are at the level where the renormalization 
must be taken into account, as was also done in several other recent large-scale studies \cite{hasenbusch05a,hasenbusch05b,hasenbusch08,komura12}. 

In addition to the multiplicative renormalization of the stiffness, a different factor has also been found in the leading logarithmic correction.
According to Hasenbusch et al.~\cite{hasenbusch05a,hasenbusch05b}, the correction in Eq.~(\ref{rhoslogcorr1}) 
should be modified to read
\begin{equation}
\frac{\rho_s(T_{\rm BKT},L)}{\rho_s(T_{\rm BKT},\infty)}=1 + \frac{g}{2\ln(L) + C+\ln[C/2 + \ln(L)]} + \frac{a}{\ln^2(L)}+\ldots,
\label{rhoslogcorr2}
\end{equation}
where  $g=1.00202783$. This constant is also very important in proper finite-size scaling studies with high-precision data.

\subsection{Finite-size scaling procedures}

The leading WM log-correction (\ref{rhoslogcorr}) has been used extensively to analyze MC data in the past. In finite-size extrapolations of $T_{\rm BKT}$ the
most common procedure has been to find the best value of $C$ to fit a series of finite-size data \cite{melko04,hasenbusch05a,hasenbusch08}. Another way is to 
divide out the factor containing the logarithm, with $C$ chosen such that curves graphed versus the temperature for different system size cross each other within 
as narrow a range of $T$ as possible (with the crossing points for large lattices approaching the BKT temperature) \cite{carrasquilla12}. With 
the log-correction divided out, curves for different system sizes graphed versus $T$ can also be scaled to collapse onto each other remarkably well 
by using the known exponential divergence of the correlation length \cite{harada97,sandvik10}.

We already outlined our alternative finite-size scaling approach in Sec.~\ref{sec:intro}. With the MC-calculated stiffness constants we want to 
satisfy Eq.~(\ref{rhol1l2}) with $\rho_s$ replaced by $\rho^{\rm MC}_s$ and using either of two different forms ($i=1,2$) of the correction factor;
\begin{equation}
\frac{\rho^{\rm MC}_s(T^*,L_1)}{F_i(L_1)} = \frac{\rho^{\rm MC}_s(T^*,L_2)}{F_i(L_2)} = f\frac{2T^*}{\pi}.
\label{rhol1l2mc}
\end{equation}
The $F$-functions correspond to the WM correction in (\ref{rhoslogcorr}) and the higher-order form in (\ref{rhoslogcorr2}), in both cases including 
the correction factor $g$:
\begin{eqnarray}
F_1(L) & = & 1 + \frac{g}{2\ln(L) + C}, \label{f1def} \\
F_2(L) & = & 1 + \frac{g}{2\ln(L) + C+\ln[C/2 + \ln(L)]}. \label{f2def}
\end{eqnarray}
In both these forms the single free parameter $C$ is adjusted to satisfy Eq.~(\ref{rhol1l2mc}) at some temperature $T^*$ for two system sizes. 
The relationship between the sizes should be arbitrary and we here use $L_1=L$ and $L_2=2L$.

\begin{figure}
\begin{center}
\includegraphics[width=9cm, clip]{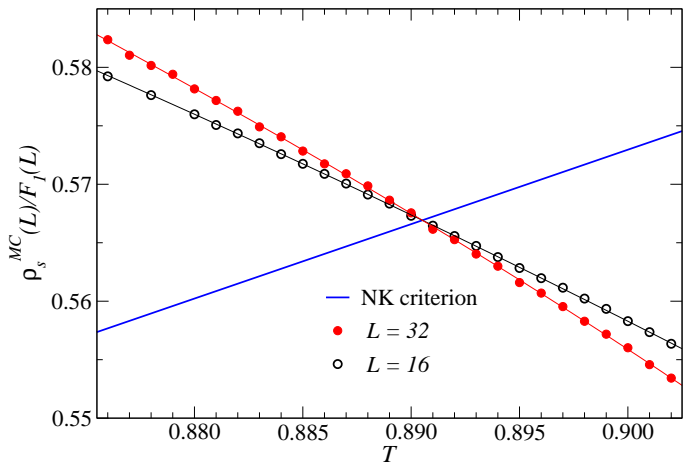}
\end{center}
\vskip-4mm
\caption{Illustration of the fitting procedure based on Eq.~(\ref{rhol1l2mc}), using the correction function $F_1$ and system sizes $L_1=16$ and $L_2=32$. 
The constant $C=C(L_1,L_2)=1.271$ in the function $F_1$ [defined in (\ref{f1def})] has has been chosen such that the two $\rho^*_s/F_1$ curves 
(in terms of polynomials fitted to the MC data points; shown here with the continuous curves) cross each other exactly at the temperature 
satisfying the NK criterion as in Eq.~(\ref{rhol1l2mc}).}
\label{cross}
\end{figure}

To illustrate the procedure, in Fig.~\ref{cross} we graph the three quantities in Eq.~(\ref{rhol1l2mc}) versus $T$ (replacing $T^*$ by $T$) in the case of 
$L=16$ and using the size-correction $F_1$. Here $C$ has been adjusted so that the curves cross each other at a common point, where the temperature $T=T^*$. 
In Sec.~\ref{sec:res} we will analyze the $L$ dependence of the crossing point as well as the behavior of the parameter $C$.

\section{Monte Carlo calculations}
\label{sec:mc}

We use standard MC methods, primarily implemented using GPU computing (as discussed below), to calculate the stiffness (helicity modulus) for the 
classical 2D XY model with Hamiltonian
\begin{equation}
H=-\sum_{\langle ij\rangle} {\vec S}_i \cdot {\vec S}_j=-\sum_{\langle ij\rangle} \cos(\Theta_i-\Theta_j),
\label{hamxy}
\end{equation}
where the spins $\vec S_i$ are 2D vectors of length $S=1$ and the expression in terms of the angles $\Theta_i$ is more convenient in practice. 
We here first discuss the definition of the helicity modulus and then outline the MC algorithms and their GPU implementation.

\subsection{The helicity modulus}

The helicity modulus is defined according to
\begin{equation}
\rho_a = \left.\frac{1}{N} \frac{\partial^2 G(\phi)}{\partial \phi^2}\right|_{\phi=0},
\label{stifffdef}
\end{equation}
where $G(\phi)$ is the free energy in the presence of a twist field (or, equivalently, a twisted boundary condition) in the lattice direction
$a$ ($a=x,y$). The MC estimator for this quantity, computed in simulations at $\phi=0$, is given by
\begin{equation}
\rho^{\rm MC}_a=\frac{1}{L^2}\left ( \langle H_a\rangle -\frac{1}{T} \langle I_a^2\rangle \right ),
\label{rhosestimator2cl}
\end{equation}
where $H_a$ is the Hamiltonian including only the $a$-directed links (nearest-neighbor site pair) in (\ref{hamxy}) and $I_a$ is the ``current'' in 
the $a$ direction, given by
\begin{equation}
I_a=-\sum_{\langle i,j\rangle_a}\sin(\Theta_j-\Theta_i).
\label{ixdef}
\end{equation}
A pedagogical derivation of these expressions can be found in Ref.~\cite{sandvik10}. 

\subsection{GPU computing}
\label{gpumc}

Here we summarize the procedures used in our MC simulations on the GPU, which we have implemented using the NVIDIA CUDA framework. 
We refer interested readers to available literature for an introduction to the details of the GPU hardware and the programming models \cite{CUDA}. 

We use parallel Metropolis single-spin flips as well as over-relaxation moves \cite{Creutz1987,Li1989}. In addition, to improve the dynamics, and 
for convenience when computing stiffness constants for a range of temperatures close to the transition, we run several temperatures simultaneously and 
apply parallel-tempering (PT) \cite{Hukushima1996}, where  configurations for nearby temperatures are occasionally swapped  (using the Metropolis 
acceptance probability). One MC step (MCS) is then defined as one Metropolis sweep, an over-relaxation sweep of the entire lattice, followed by 
one parallel-tempering exchange attempt for each pair of adjacent temperatures. 

The over-relaxation algorithm was used by Gupta et al. \cite{gupta88} for the same model as we study here and by Wolff for a
different model \cite{wolff92}. The optimal ratio of Metropolis to over-relaxation updates was discussed in these works.
Here we are not studying very large lattices and we did not optimize the ratio as a function of temperature
and lattice size. We simply use a mix which leads to comparable times spent on single-spin and relaxation updates (and the time taken by the parallel 
tempering is negligible).

\begin{figure}[tb]
\begin{center}
\includegraphics[angle=90,width=0.45\textwidth,clip]{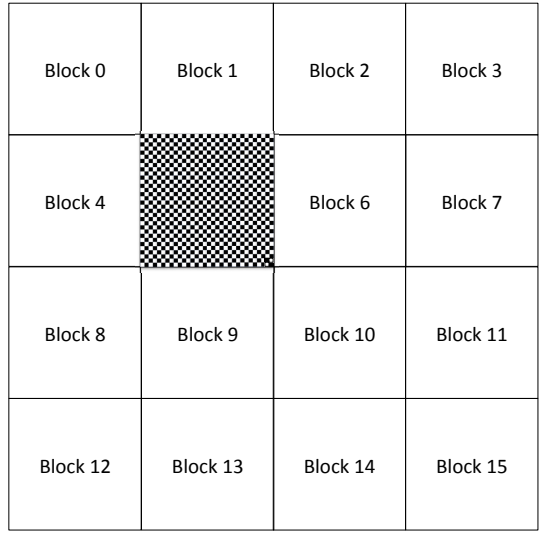}
\end{center}
\vskip-4mm
\caption{Mapping of a $128\times 128$ lattice to thread blocks on GPU. Each thread block of $16 \times 16=256$ threads 
performs MC updates on $32 \times 32=1024$ spins.}
\label{fig:2Dblocks}
\end{figure}

To implement the parallel Metropolis and over-relaxation updates in a way suitable for the GPU, we divide the entire lattice into blocks of  
$32 \times 32 = 1024$ spins. Each block is  decomposed into  two different sub-lattices, as shown in Fig.~\ref{fig:2Dblocks}. 
Each block is assigned to a \textit{thread block} \cite{CUDA} containing $16\times 16=256$ threads, which execute the same GPU \textit{kernel} in 
parallel \cite{CUDA}.  Each thread is responsible for updating $2\times 2=4$ spins,  with two ``black'' sites and two ``white'' sites, so that there 
are enough arithmetic operations to hide the latency of the global memory accesses \cite{CUDA}. We  apply the  checkerboard decomposition algorithm 
to perform the Metropolis single-spin flips in parallel \cite{Preis2009,Weigel2011}. We first update all the black sublattce spins in parallel via a 
GPU kernel. After all the black spins belonging to different blocks are updated,  another kernel is launched to  update all the white sublattice spins. 

Due to the special 
architecture of the GPU,  the commonly used Mersenne-Twister (MT) random number generator can not be efficiently implemented at the thread level. 
Instead,  we use a faster generator especially designed for the GPU architecture; the \textit{Warp Generator}~\cite{WarpGenerator}. We note 
that although it has a smaller period of $2^{1024}-1$ than the MT ($2^{19937}-1$), this period still far exceeds the length of the sequence used
in practice. We also do not find any noticeable differences between results when compared with conventional CPU runs using the MT generator.

It is well established that the single-spin flip  Metropolis update suffers from critical slowing down near phase transitions and for 
increased efficiency one has to resort to cluster updates \cite{Swendsen1987, Wolff1989}. However, GPU implementations of the cluster update 
are complicated and less efficient \cite{Weigel2011b}. We instead implemented the microcanonical over-relaxation update \cite{Creutz1987,Li1989}
and found it to be as efficient as the cluster update in reducing slowing-down. It should also be noted that slowing-down is not very serious
at the BKT transition compared to standard critical points. 

In an over-relaxation move, the new spin direction on site $i$ is obtained by reflecting it with respect to its 
local molecular field, 
\begin{equation}
\mathbf{H}_i=- \sum_{\langle ij\rangle} \mathbf{S}_j, 
\end{equation}
according to
\begin{equation}
\mathbf{S}_i'=-\mathbf{S}_i+2\frac{\mathbf{S}_i \cdot \mathbf{H}_i}{H_i^2}\mathbf{H}_i.
\end{equation}
This update maps the system from a point in the phase space to another point with exactly the same energy. After several sweeps, the system 
is able to explore a larger region of the phase space without being stuck at a particular local minimum for a long time, thus improving the 
ergodicity of the simulation. 

To better equilibrate the simulations and further reduce slowing-down effects close to the transition, we also perform PT sweeps \cite{Hukushima1996}
on many systems at different temperatures simulated simultaneously. After a certain number of MCSs (typically just one), we swap two adjacent configurations 
$X_m, X_n$ at neighboring temperatures $T_m, T_n$  with the acceptance probability of
\begin{equation}
W(X_m, T_m | X_n, T_n) = {\rm min} \left[ 1,e^{(1/T_m - 1/T_n)(E_m - E_n)} \right],
\end{equation}
where  $E_n$ is the total energy of replica $n$.

To reduce the amount of data transfer between the CPU and the GPU, we store all the spin configurations at different temperatures in the GPU global 
memory, and all updates are performed through the kernel functions on the GPU. Measurements are also performed on the GPU and the results are sent back 
to the CPU for data binning. Simulations were carried out at 21 temperatures ranging from $T = 0.888$ to $T = 0.898$ for system sizes ranging from 
$L = 16$ to $L = 512$ in steps of 16 (to keep optimal sizes for the GPU memory structure, as illustrated in Fig.~\ref{fig:2Dblocks}). In each simulation, 
about $10^8$ measurements were made after $ 10^6$ MCSs for equilibration. The data were blocked into bins of $10^5$ measurements, which were subject to 
further statistical analyses post-simulation. The simulations were performed on Tesla C2090 GPUs, and took approximately 3600 GPU hours for producing 
the whole data set discussed in this paper.

We also used standard CPUs with single-spin and cluster updates for small systems. For the range of systems where we have results
from both CPU and GPU calculations, they agree perfectly within statistical errors.

\section{Results}
\label{sec:res}

We here use system pairs of the form $(L,2L)$ and extract crossing points such as the one shown in Fig.~\ref{cross}. Note again
that the parameter $C$ depends on $L$, and for large $L$ we expect different behaviors depending on which one of the size-corrections, 
Eq.~(\ref{f1def}) or (\ref{f2def}), is used. Comparing the two forms, we see that $F_2$ can be reproduced by $F_1$ if $C$ is of the form 
$C = C_0 + \ln [C_0/2+\ln(L)]$ in the latter. When using $F_2$, $C$ should converge to a constant for large $L$, unless there are further 
higher-order logarithms in the denominator. A divergent $C$ could also in principle result from other logarithmic corrections that
can be mimicked by the function $F_2$.

Beyond corrections that can be effectively included in $F_1$ and $F_2$ through the single parameter $C$, there are also other corrections, 
as discussed in Sec.~\ref{sec:corr}. The size-dependent transition temperature $T^*(L)$ is extracted in a rather convoluted way and it is not 
{\it a priori} clear exactly how the corrections in $\rho_s$ translate into an $L$-dependence of $T^*(L)$. One may, nevertheless, expect the 
general form of the correction in $\rho_s$ to survive in $T^*$, i.e., there should be logarithmic corrections of the form $1/\ln^2(L)$ as 
in Eq.~(\ref{rhoslogcorr2}). We will test different forms of corrections to investigate the sensitivity of the final extrapolated $T_{\rm BKT}$.

In this section we first discuss a few more details of the procedures used to extract $T^*(L)$ and then study the convergence properties 
of the transition temperature and the behavior of the constant $C$.

\subsection{Extracting NK crossing points}

To systematically carry out the analysis illustrated in Fig.~\ref{cross}, we fit a polynomial (typically of second or third order) 
to a range of MC data for the two system sizes close to the transition. The crossing point corresponding to the first equality in
Eq.~(\ref{rhol1l2mc}) is extracted using the polynomials. The deviation from the desired NK value (the second equality) is then
minimized (to zero within machine precision) using bisection. Error bars are computed by repeating this procedure for a large number 
(hundreds) of bootstrap samples of the data. 

\subsection{Using the leading size-correction $F_1(L)$}

\begin{figure}
\begin{center}
\includegraphics[width=9cm, clip]{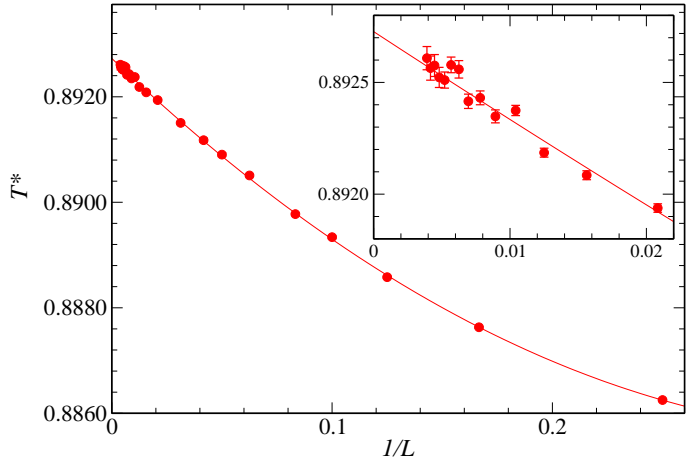}
\end{center}
\vskip-4mm
\caption{Finite-size transition temperatures extracted on the basis of system-size pairs $(L,2L)$ versus $1/L$. 
The curve is a second-order polynomial fitted to all the data points, giving the infinite-size extrapolated value of 
the transition temperature $T_{\rm BKT}=0.89273$. The inset shows the large-size data on a more detailed scale.}
\label{tktpoly}
\end{figure}

Fig.~\ref{tktpoly} shows our results for $T^*$ based on the leading-order WM form $F_1$, Eq.~(\ref{f1def}), for $L$ in the range $4$ to $256$ 
(i.e., the largest system used was $2L=512$). Although we may suspect that there should be logarithmic size corrections, it is instructive to
begin by just considering regular low-order polynomial fits to the data. A second-order polynomial in fact gives a statistically acceptable
fit to all the data starting with $L$ as small as $4$, and removing small system sizes does not significantly affect the extrapolated $L\to \infty$
value. The size dependence is weak and essentially linear, with a very small quadratic correction required when including small sizes. 
Naturally, the standard deviation of the extrapolated result increases as the data set becomes smaller. For example, including 
all the data points starting from $L=4$ we obtain $T_{\rm BKT}=0.89273(1)$, where the number within parenthesis is the standard deviation 
of the preceding digit. Starting instead from $L=32$ we obtain $T_{\rm BKT}=0.89276(3)$. These numbers agree within statistical errors 
and the statistical quality of the fit is reasonably good and similar in the two cases.

To our knowledge, the best previous result for $T_{\rm BKT}$ of the 2D XY model, obtained recently in a large-scale GPU study \cite{komura12} 
with system sizes up to $L=65535$ (based on studying relaxation dynamics starting from a high-temperature state) was $T_{\rm BKT}=0.89289(6)$, 
which deviates from our result by about $2.5$ standard deviations. i.e., the calculations are marginally consistent with each other. A similar result, 
$T_{\rm BKT}=0.8929(1)$ [actually quoted as $1/T_{\rm BKT}=1.1200(1)$] was obtained in Ref.~\cite{hasenbusch05a}. 
Given the reasonably good agreement with the previous results, one might conclude that the NK crossing procedure avoids logarithmic corrections 
through the variability of $C$. However, such a conclusion is premature, as it is hard to see how all higher-order logarithmic corrections could 
have been completely eliminated (or why corrections of the plolynomial form should appear at all). We therefore proceed to study fitting functions 
of the expected logarithmic forms.

\begin{figure}
\begin{center}
\includegraphics[width=9cm, clip]{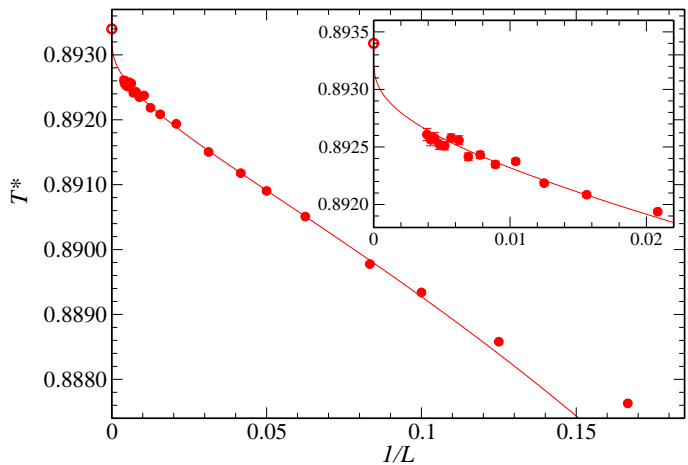}
\end{center}
\vskip-4mm
\caption{The same data as in Fig.~\ref{tktpoly} but with a fit of the logarithmic form (\ref{logfit1}), which extrapolates
to $T_{\rm BKT}=0.8934$ in the thermodynamic limit (shown as the open circle at $1/L=0$). System sizes $L \ge 12$ were included in the fit.
The inset is a more detailed plot for the largest systems.}
\label{tktlog1}
\end{figure}

\begin{figure}
\begin{center}
\includegraphics[width=9cm, clip]{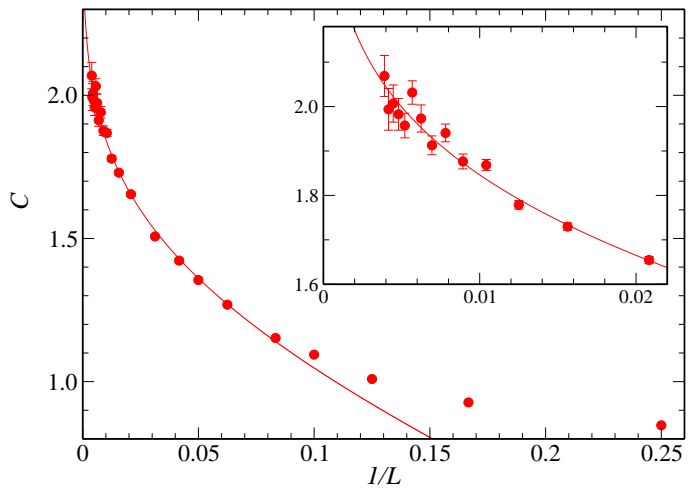}
\end{center}
\vskip-4mm
\caption{Size dependence of the constant $C$ in the WM logarithmic correction (\ref{f1def}). The curve is a fit
to the log-divergent form (\ref{cfit}). The inset shows the data for the larger systems on a more detailed scale.}
\label{clog1}
\end{figure}

Motivated by the discussion in Sec.~\ref{sec:corr}, we test the following forms
\begin{eqnarray}
T^*(L) &=& T_{\rm BKT}(\infty)+\frac{a}{\ln^2(bL)},\label{logfit1} \\
T^*(L) &=& T_{\rm BKT}(\infty)+\frac{a\ln[\ln(cL)]}{\ln^2(bL)}. \label{logfit2}
\end{eqnarray}
Interestingly, the two forms both work very well and produce almost identical (visually indistinguishable) fits with the same
extrapolated $T_{\rm BKT}=0.89340(5)$. The constant $c$ in Eq.~(\ref{logfit2}) comes out very close to $0$ (of the order $10^{-20}$ or
smaller) and therefore the form effectively reduces to the same as Eq.~(\ref{logfit1}) for our moderate sizes $L$. The latter
fit is shown in Fig.~\ref{tktlog1} (and the former one looks identical on the scale of the graph). The statistical quality of
the fit in this case is good, actually somewhat better than the polynomial fit in Fig.~(\ref{tktpoly}), which can even be seen 
by visual inspection of the large-size data in the two figures.

The fact that Eqs.~(\ref{logfit1}) and (\ref{logfit2}) produce essentially identical fits also indicates that the variable $C$ 
in $F_1$ actually reproduces the more complicated logarithmic form in $F_2$, i.e., by comparing the two we should have $C=C_0+\ln[C_0/2+\ln(L)]$.
There is an ambiguity here, however, since our $C=C(L,2L)$ is extracted based on two sizes, and it is not clear which of the two sizes
(if any) should be used in the fit to data graphed versus the smaller size $L$. In principle $C$ may also account for some of the
corrections of the form $1/\ln^2(L)$ in (\ref{rhoslogcorr2}), which would also case deviations from the above form. To account for
the uncertainty associated with the system size, we introduce another parameter, $\lambda$, fitting to the form 
\begin{equation}
C=C_0+\ln[C_0/2+\ln(L/\lambda)]. 
\label{cfit}
\end{equation}
This fit works remarkably well for $L>10$, as shown in Fig.~\ref{clog1}. The constant $\lambda \approx 2$ when all the data starting from 
$L=12$ are used, but when removing several small sizes $\lambda \approx 1$ also works well. The growth of $C$ with $L$ has also been noticed in 
previous works \cite{hasenbusch05a,hasenbusch08}, which monitored how the parameter changes as smaller system sizes were eliminated in a fit 
including many system sizes with a common $C$. To our knowledge the size dependence has not previously been studied in detail.

\subsection{Using the higher-order size-correction $F_2(L)$}

\begin{figure}
\begin{center}
\includegraphics[width=9cm, clip]{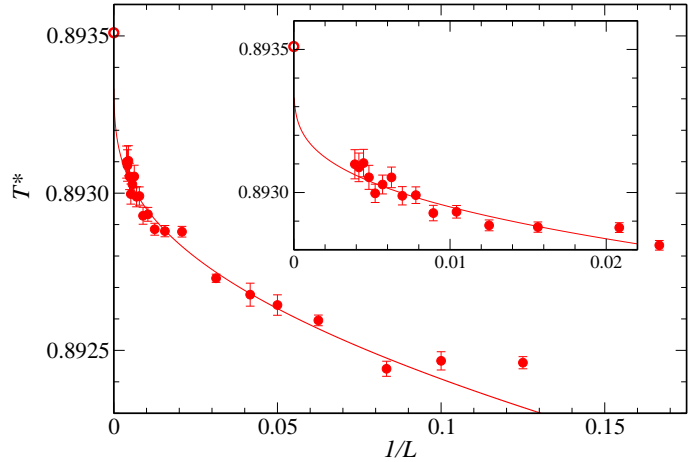}
\end{center}
\vskip-4mm
\caption{Size dependence of the transition temperature extracted using Eq.~(\ref{rhol1l2mc}) with the high-order logarithmic correction (\ref{f2def}).
The curve is a fit to the logarithmic form (\ref{logfit1}), using only $L \ge 12$ data. The extrapolation gives $T_{\rm BKT}=0.89351$ 
in the thermodynamic limit (shown with the open circle). The inset shows the large-size data on a more detailed scale.}
\label{tktlog2}
\end{figure}

We now repeat the same kind of analysis as above but with the function $F_2$, Eq.~(\ref{f2def}), used in the $T^*$ condition (\ref{rhol1l2mc}). 
Fig.~\ref{tktlog2} shows the results along with a fit to the form (\ref{logfit1}), which now is the expected correction. This form again works very well 
when used with the $L \ge 12$ data and then extrapolates to $T_{\rm BKT}=0.8935(1)$, which is consistent within error bars with the previous result when 
$F_1$ was used. The error bar here is twice as large as the one obtained with $F_1$, however. Eliminating smaller sizes, the error bar grows but 
the extrapolated value stays consistent within those error bars with the result quited above. Interestingly, in this case the $T^*$ results show even 
without extrapolations that the previous estimates of $T_{\rm BKT} \approx 0.8929$ are too low, as data for several of the largest system sizes are 
already above this value and the upward {\it nonlinear} trend is very distinct; more so than in the $F_1$-based Fig.~\ref{tktlog1}.

\begin{figure}
\begin{center}
\includegraphics[width=9cm, clip]{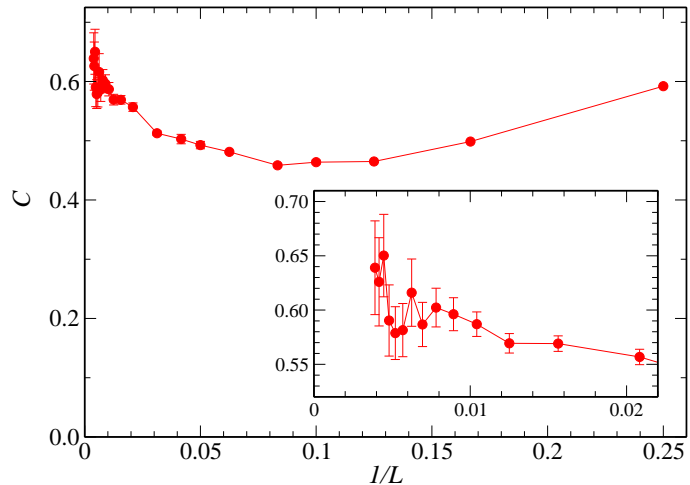}
\end{center}
\vskip-4mm
\caption{Size dependence of the parameter $C$ in the logarithmic correction (\ref{f2def}). The inset shows
the large-size data on a more detailed scale.} 
\label{clog2}
\end{figure}

The size dependence of $C$ is shown in Fig.~\ref{clog2}. Here it is not possible to conclude whether there is convergence to a constant when 
$L \to \infty$ or whether there is some very weak logarithmic divergence left. One can certainly make good fits to functions of either kind. It is in any 
case clear that the size dependence is much weaker than in the $F_1$-based data shown in Fig.~\ref{clog1}.

\section{Summary}
\label{sec:disc}

We have presented an improved finite-size scaling method for studying the BKT transition. Taking advantage of the NK relationship (\ref{rhojump})
governing the spin stiffness at the transition temperature $T_{\rm BKT}$ in the thermodynamic limit, we defined a two-size estimate (using a curve-crossing 
criterion) for the transition temperature which is constrained by this relationship also for finite size. We tested the procedure for the standard 2D XY model, 
with high-precision finite-size data obtained by MC simulations on GPUs, for lattice sizes up to $512\times 512$. 

We used two forms of the logarithmic correction to the size-dependence of the spin stiffness at the transition point, Eq.~(\ref{rhotstar}). 
The first one, Eq.~(\ref{f1def}), is essentially the long-known WM form \cite{weber87}, while the second one, Eq.~(\ref{f2def}), is a more
recently derived higher-order form \cite{hasenbusch05a,hasenbusch08,pelissetto13}. The key to our approach is that both of these forms contain a 
single adjustable parameter $C$, which together with the NK relation enables a unique definition of the two-size transition temperature $T^*$. 

With regards to previous studies using the WM form with a common $C$ fitted to all system sizes, it is important to note that in our approach we have 
shown that the constant $C$ diverges when the WM form is used (as it should based on theoretical expectations \cite{hasenbusch05a,hasenbusch08}). Thus, 
any approach based on fitting data for a range of system sizes to the WM form with a common value of $C$ is strictly speaking incorrect. The
effect of changing $C$ becomes unimportant only for system sizes larger than what can be studied in practice. Moreover, the size dependence of $C$ 
effectively can account for some, but not all, of the higher-order corrections.

\begin{figure}
\begin{center}
\includegraphics[width=9cm, clip]{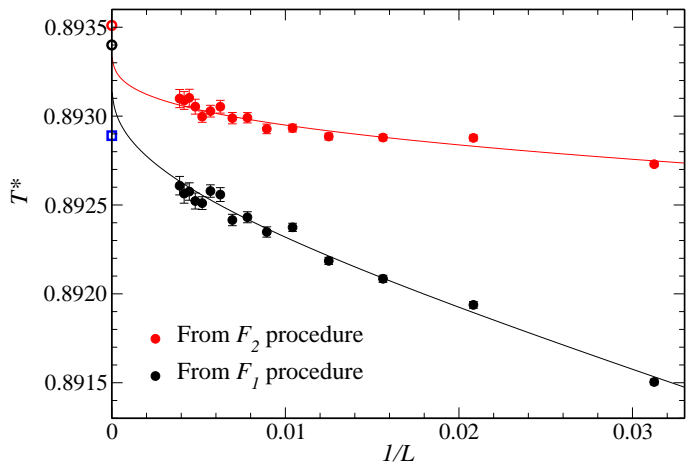}
\end{center}
\vskip-4mm
\caption{Comparison of transition temperatures extracted using the log-corrections of type $F_1$ and $F_2$ (solid circles). 
The data and fitted curves are the same as those in Figs.~\ref{tktlog1} and \ref{tktlog2}, with the circles showing the corresponding
extrapolations to infinite size. The solid square is the result obtained by Komura et al.~\cite{komura12}.}
\label{tktlog3}
\end{figure}

It is instructive to compare directly the remaining size dependence of $T^*$ and its approach to $T_{\rm BKT}$ when the two different forms 
of the log corrections are used. A graph with only data for the larger lattices is shown in Fig.~\ref{tktlog3}. It is clear that the more sophisticated 
form $F_2$ leads to a significantly faster convergence, but it should be noted that the method is unbiased in both cases. It is, however, crucial to 
extrapolate the results using the expected logarithmic corrections beyond those in the $F$-functions. The form (\ref{logfit1}) works very well in both 
cases and the fits demonstrate that results of previous high-precision studies most likely were affected by the neglect of these corrections.
Especially with $F_2$-scheme, the raw data already are above the previous results and the further upward trend is clear. Based on our work,
we present $T_{\rm BKT}=0.8935(1)$ as our best estimate of the transition temperature.

\begin{figure}
\begin{center}
\includegraphics[width=9cm, clip]{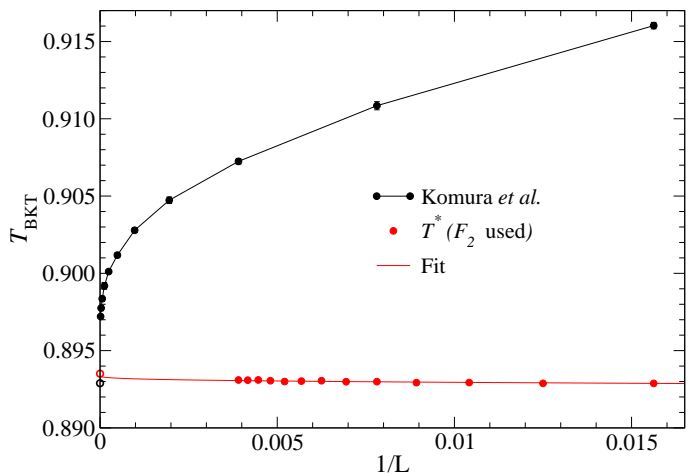}
\end{center}
\vskip-4mm
\caption{Comparison of our finite-size estimates of $T_{\rm BKT}(L)=T^*$ (with the logarithmic fit shown as the solid curve) and those of 
Ref.~\cite{komura12} (where the lines between data points only provide a guide to the eye).}
\label{compjap}
\end{figure}

It is also interesting to compare our finite-size data directly with those of Komura et al.~\cite{komura12}, which we do in Fig.~\ref{compjap}. 
Note that the finite-size definitions of $T_{\rm BKT}$ are very different in these two calculations, and that the data of Ref.~\cite{komura12} were 
originally analyzed in a different way, with only a leading logarithmic correction (adjusted for the best fit to the data) to the infinite-size 
$T_{\rm BKT}$. The comparison is still very illuminating. It is clear that our definition of the transition temperature has much smaller size corrections. 
Moreover, since we have shown here that the sub-leading logarithmic corrections are important, it is likely that the leading-log extrapolations 
in Ref.~\cite{komura12} underestimate the transition temperature, even with the very large systems used. Such results may mimick the polynomial 
fits used in Fig.~\ref{tktpoly}. 

Besides the spin stiffness that we have studied here, it is possible to use the method with other dimensionless quantities as well, e.g., 
the  Binder cumulant and the ratio $\xi/L$, $\xi$ being the correlation length. All these quantities were previously analyzed by Hasenbusch 
et al.~\cite{hasenbusch08} in a way resembling our treatment with the correction factor $F_1$, but no systematic studies or extrapolations 
of $T_{BKT}$ or $C$ were carried out. 

While our method is unbiased and works well with the correction factor of either type $F_1$ or $F_2$, Eqs.~(\ref{f1def}), (\ref{f2def}), we recommend 
the latter because the remaining finite-size corrections are significantly smaller. In either case it is crucial to also extrapolate the final result 
using the next known logarithmic correction (\ref{logfit1}). The method should be generally applicable to a wide range of BKT transitions and the 
scheme is rather simple to implement in practice.

It would be interesting to study the standard XY model to even higher precision on larger lattices, especially to investigate 
further the asymptotic large-$L$ behavior of the constant $C$ when the $F_2$-scheme is used. It is presently not clear whether it
converges or diverges, although the results shown in Fig.~\ref{clog2} certainly indicate that the size dependence is very weak for
large systems. Given the rather modest GPU resources we have used in the present work, it will certainly be possible to go to
considerably larger sizes in the near future.

When studying other, more complicated models exhibiting BKT transitions (including quantum models) one can use a simpler model such as the standard XY model 
as a point of reference in a ``matching method'' \cite{hasenbusch97,ceccarelli13}. High-precision results, including a good estimate of $T_{\rm BKT}$, for the 
reference model are needed to make this approach unbiased. This application also motivates further high-precision GPU MC studies of the 2D XY model. It should 
also be possible to adapt the matching method to the combined NK and curve-crossing approach we have discussed in this paper.

\vskip7mm

\noindent{\bf Acknowledgments}
\vskip3mm

\noindent
We would like to thank Martin Hasenbusch and Ettore Vicari for several very useful discussions about the form of
the higher-order logarithmic corrections and their importance.
This work was supported by NSC in Taiwan through Grant No.~100-2112-M-002-013-MY3 (YDH and YJK), NTU Grant number 101R891004 
(YJK), and by the NSF under Grants No.~DMR-1104708 and PHY-1211284 (AWS). AWS also gratefully acknowledges support from the NCTS 
in Taipei for visits to National Taiwan University.

\end{document}